\documentclass[11pt,twoside]{amsart}
\pdfoutput=1

\usepackage{amsmath}
\usepackage{amsxtra}
\usepackage{amscd}
\usepackage{amsfonts}
\usepackage{amssymb}
\usepackage{eucal}
\usepackage{epsf}
\usepackage{cite}
\usepackage{a4}
\usepackage{microtype}

\let\a=\alpha \let\be=\beta  
 \let\z=\zeta \let\h=\eta 

 \let\k=\kappa \let\la=\lambda 
\let\n=\nu \let\x=\xi \let\p=\pi \let\r=\rho \let\s=\sigma
\let\om=\omega 
  \let\PH=\Phi 
  
  \let\D=\Delta

\let\qd=\quad  

\def\epp{\, .}
\def\epc{\, ,}

\def\2{\frac{1}{2}} \def\4{\frac{1}{4}}

\def\6{\partial}

\def\+{\dagger}

\def\<{\langle} \def\>{\rangle}

\def\rd{{\rm d}}

 \DeclareMathOperator{\cth}{cth}

\def\Re{{\rm Re\,}}

\def\fa{\mathfrak{a}}

\renewcommand{\tilde}{\widetilde}

\newcommand{\betab}{\mbox{\boldmath$\beta$}}
\newcommand{\gammab}{\mbox{\boldmath$\gamma$}}

\newcommand{\rhob}{\rho^\mathrm{sc}}
\newcommand{\omegab}{\omega^\mathrm{sc}}

\newcommand{\taub}{\mbox{\boldmath$\tau$}}

\newcommand{\cb}{\mathbf{c}}
\newcommand{\cbs}{\mathbf{c^*}}
\newcommand{\bb}{\mathbf{b}}
\newcommand{\bbs}{\mathbf{b^*}}

\newcommand{\tb}{\mathbf{t}}

\newcommand{\nn}{\nonumber}
\newcommand{\bea}{\begin{eqnarray}}
\newcommand{\ena}{\end{eqnarray}}
\def\bel{\begin{eqnarray}}
\def\enl{\end{eqnarray}}
\newcommand{\en}{\end{eqnarray*}}
\newcommand{\ba}{\begin{array}}
\newcommand{\ea}{\end{array}}

\newcommand{\slth}{\widehat{\mathfrak{sl}}_2}

\newcommand{\Tr}{{\rm Tr}}

\newenvironment{tenumerate}{
  \begin{enumerate}
  
  }{\end{enumerate}}
\newcommand{\bi}{\begin{tenumerate}}
\newcommand{\ei}{\end{tenumerate}}
\newcommand{\isoto}[1][]%
{{\mathop{\buildrel{\sim}\over\longrightarrow}\limits_{#1}}}

\def\[{\left[}
\def\]{\right]}
\newcommand{\al}{\alpha}

\numberwithin{equation}{section}
\newtheorem{thm}{Theorem}[section]

\newtheorem{lem}[thm]{Lemma}

\theoremstyle{definition}

\newtheorem{rem}[thm]{Remark}
\newtheorem*{ack}{Acknowledgments}
\newtheorem*{rem*}{Remark}

\newcommand{\bts}{\mathbf{t^*}}

\newcommand{\bc}{\mathbf{c}}
\newcommand{\lb}{\mathbf{l}}

\def\bi{\mathbf{i}}

\begin{document}

\begin{title}[
]{Properties of linear integral equations related to the six-vertex
model with disorder parameter II}
\end{title}

\date{\today}

\author{Hermann Boos$^{\dag\ddag}$ and Frank G\"ohmann$^\dag$}


\address{$^\dag$~Fachbereich C -- Physik, Bergische Universit\"at
Wuppertal, 42097 Wuppertal, Germany}

\address{$^\ddag$~Skobeltsyn Institute of Nuclear Physics, Moscow
State University, 119991 Moscow, Russia}

\email{boos@physik.uni-wuppertal.de,
goehmann@physik.uni-wuppertal.de}


\begin{abstract}
We study certain functions arising in the context of the calculation
of correlation functions of the XXZ spin chain and of integrable
field theories related with various scaling limits of the underlying
six-vertex model. We show that several of these functions that are
related to linear integral equations can be obtained by acting with
(deformed) difference operators on a master function $\Phi$. The
latter is defined in terms of a functional equation and of its
asymptotic behavior. Concentrating on the so-called temperature case
we show that these conditions uniquely determine the
high-temperature series expansions of the master function. This
provides an efficient calculation scheme for the high-temperature
expansions of the derived functions as well.
\end{abstract}

\maketitle

\section{Introduction}
In \cite{BJMST06b,BJMST08a,JMS08,BJMS09a} an algebraic structure of
the correlation functions of the XXZ model (or of the underlying six-vertex
model) called factorization was identified in a rather general setting. It
was shown that the correlation functions consist of an algebraic part
and of a physical part. The algebraic part is determined by a set of
operators $\bts$, $\bb$, $\bc$, $\bbs$, $\cbs$ which act on a space
${\mathcal W}^{(\alpha)}$ of quasi-local operators. They do not
depend on the physical parameters like temperature, magnetic field
etc. The operators $\bb$, $\bc$ (annihilation operators) and $\bbs$,
$\cbs$ (creation operators) are two pairs of Fermi operators
satisfying characteristic anti-commutation relations. The creation
operator $\bts$ is `bosonic' and commutes with all Fermi operators.
With the help of the creation operators one can construct a
`fermionic basis' of the space of quasi-local operators. In contrast to the
algebraic part the physical part depends on the physical parameters
and is represented by two transcendental functions $\rho$ and $\omega$.

The proof of the factorization in \cite{JMS08} covers the rather general
case of a six-vertex model which is finite in vertical, so-called
Matsubara direction and carries arbitrary inhomogeneity parameters
on the horizontal lines. The proof is based on the properties of the
fermionic and bosonic operators. It also relies on a generalization of
the density matrix (denoted $Z^\kappa$ in \cite{JMS08}), such as to include
a disorder parameter $\alpha$, and on the use of so-called $q$-deformed
Abelian integrals of the second kind. The transcendental functions $\rho$
and $\omega$ which constitute the physical part also depend on one or
two spectral parameters $\z,\z'$. The function $\rho (\z)$ is the ratio
of two maximal transfer matrix eigenvalues with different twist parameters.
It is related to the one-point correlator. The function $\om (\z, \z')$
is related to the nearest-neighbor correlators, but its definition is
more involved. The authors of \cite{JMS08} demonstrated that it is determined
by a `normalization condition' for their $q$-deformed Abelian integrals
of the second kind.

In \cite{BoGo09} we suggested an alternative description of the function
$\om$ by means of the solutions of certain linear and non-linear
integral equations. We also proved the equivalence of this description
and the one suggested in \cite{JMS08}. In \cite{BoGo10} we discussed some
properties of those integral equations in the case of finite temperature.
The normalization condition for $\om$ turned out to be a consequence of
the integral equations for the auxiliary functions $\fa$ and $G$ introduced
in \cite{Kluemper92,Kluemper93,GKS04a,BoGo09}. An important point in
\cite{BoGo10} was to introduce the `dressed charge' $\sigma$ which, like
the function $G$, fulfills a linear integral equation. Using Baxter's
TQ-relation, it appeared to be possible to find a solution to this
equation in `explicit form' in terms of the ratio $\Phi$ of eigenvalues of
two $Q$-operators with different twist parameters.

In our recent works \cite{BJMS10,Boos11} we studied the scaling
limit towards CFT on a cylinder. We conjectured that the creation
operators $\bts$, $\bbs$, $\cbs$ have well-defined scaling limits
$\taub^*$, $\betab^*$, $\gammab^*$. We could identify the states
generated by their action on the vacuum with the Virasoro module
of the CFT descendant states up to level 8. This identification was achieved
by considering the three-point CFT correlators on the cylinder with a
descendant field $P_\alpha (\{\lb_{-n}\}) \phi_\alpha (0)$ at the
origin and two asymptotic `edge'-fields $\phi_{\kappa}(\infty)$ and
$\phi_{-\kappa'}(-\infty)$ at $\pm \infty$ or their descendants,
where the primary field $\phi_{\al}$ has conformal dimension
$\Delta_\alpha = \nu^2 \alpha (\alpha - 2) /(4 (1 - \nu))$. On the
other hand we took the scaling limit of the determinant formula
derived in \cite{JMS08} for the generating function of correlation
functions of the XXZ model, which is realized as a six-vertex model
on a lattice with cylindrical topology and with creation operators
inserted. This involves, in particular, taking the scaling limits
$\rhob$ and $\omegab$ of $\rho$ and $\omega$. Comparing the
coefficients of the asymptotic expansions at large spectral
parameter, we could perform the above identification. Unfortunately,
this was possible only in a weak sense, i.e.\ modulo integrals of motion.

Let us further comment on the latter point, because it is of particular
importance for the motivation of the present paper. First, we considered
the six-vertex model at finite disorder parameter $\alpha$ and magnetic
field $\kappa$ which also plays the role of a twist or flux parameter.
In fact, in order to determine the correlation functions on the lattice
in such a way that they have appropriate scaling limits, we also needed to
introduce a kind of `lattice screening operators' that change the spin
by some number $s$. Then we came to a picture in which the twist parameter
on the `left' edge of the cylinder is $\kappa$ and on the `right' edge
$\kappa' = \kappa + \alpha + 2 s (1 - \nu)/\nu$, where $\nu$ is related
to the deformation parameter $q=e^{\pi \nu i}$. We argued that $\kappa'$
becomes an independent variable in the scaling limit. Thus, expectation
values of quasi-local operators turn into three-point functions of two
independent primary fields at $\pm \infty$ and a descendent field at $0$
as was pointed out above. When $\kappa'=\kappa$ it follows that $\rho=\rhob=1$,
and the integrals of motion do not contribute. In this case we managed
to develop (see \cite{BJMS10,Boos11}) a technique for the calculation
of the coefficients in the asymptotic expansion of the function $\omegab$
based on Wiener-Hopf factorization.

The problem is how to extend this to the general case $\kappa'\ne\kappa$
and $\rho\ne 1$. It seems we have to develop a new technique here. This
work is a first attempt in this direction. We consider the temperature
case, where the high-temperature expansion is a powerful tool to
generate explicit results.

We shall take up some ideas of the paper \cite{BoGo10}, where we studied
the properties of linear integral equations, like the one for the dressed
charge $\sigma$, using the above mentioned function $\Phi$ in the temperature
case. The idea elaborated below is to introduce a generalized function $\Phi$
of two arguments in such a way that it satisfies a functional relation which
involves the original function $\Phi$ of one argument. Inserting the high-%
temperature expansions of the two functions we see that the expansion
coefficients of the new function are recursively determined by those of the
old function. This provides an efficient scheme to calculate them on a
computer.

We also consider the calculation schemes for the coefficients in the
high-temperature expansions of the resolvent $R$, the function $G$, its dual
$\overline G$ and the function of our main interest $\omega$. We find that they
are all connected to our new function $\Phi$ in a simple way:
\begin{align*}
     & R(\z,\z') = -\frac1{2\pi i}\Delta_{\z}\Delta_{\z'}\Phi(\z,\z') \epc \quad
       \frac14\om(\z,\z') = H_{\z}H_{\z'}\Phi(\z,\z') \epc \\
     & G(\z,\z') = \Delta_{\z}H_{\z'}\Phi(\z,\z') \epc \quad
       \overline G(\z,\z')=-\frac1{2\pi i}H_{\z}\Delta_{\z'}\Phi(\z,\z') \epc
\end{align*}
where $\Delta$ and $H$ are certain difference operators defined in (\ref{deltas}),
(\ref{H}). These formulae are valid in general, in particular, in the temperature
case and in the scaling limit. This is the main result of this paper. The function
$\Phi(\z,\z')$ has simpler analytic properties than the resolvent or the
$\om$-function and together with the function $\rho$ contains the whole information
that we need for the correlation functions. In this paper we show how to calculate
the high-temperature expansion of $\Phi(\z,\z')$. For now this seems to be
easier as compared to the problem of calculating the asymptotic expansions
in the scaling limit. We do not need to use the Wiener-Hopf technique here. Still,
we plan to obtain the asymptotic expansion of the $\Phi$-function in the scaling
limit in our future work. Although our primary aim are CFTs this may turn out
to be useful in conjunction with the recent progress in the understanding of
the massive integrable Sine-Gordon theory \cite{JMS10,JMS11pp,JMS11b,JMS11} as well.

The paper is organized as follows. In Section 2 we recall some definitions from
the papers \cite{BoGo09,BoGo10,BJMS10} and introduce those objects that we
need for the further considerations. We also discuss the basic thermodynamic
functions and the integral equations they fulfill. In Section 3 we define the
function $\Phi(\z)$ of one spectral parameter and recall how the `dressed charge'
$\sigma$ can be expressed in terms of this function. We derive an equation
for the coefficients of the high-temperature expansion of $\Phi(\z)$ which
can be solved by iteration. In Section 4 we generalize this method to solve
the linear equation for the resolvent $R(\z,\z')$. To this end we introduce
a function $\Phi(\z,\z')$ which is a generalization of $\Phi(\z)$. We
constitute Lemma 4.1 on the representation of the resolvent in terms of the
function $\Phi(\z,\z')$ and obtain its high-temperature expansion. In Section 5 we
formulate Lemma 5.1 comprising the above formulae representing the functions $G$,
$\overline G$ and $\om$ by the action of difference operators on $\Phi(\z,\z')$.
We also argue that these formulae are valid in general and not only in the
temperature case. In the Appendices we prove Lemma \ref{lem:R} and \ref{lem:omphi}
and show several lowest order coefficients of the high-temperature expansions
of the above functions.

\section{Basic objects and equations}
\subsection{Correlation functions and the functional $Z^{\kappa,s}$}
This work is about the properties of special functions arising in the
context of the calculation of correlation functions of the XXZ chain
and of conformal field theories. Before defining these functions we would
like to briefly sketch the context. For more details the reader is
referred to \cite{BJMST08a,JMS08,BJMS10}.

A correlation function is an expectation value of a local operator $\mathcal{O}$,
typically calculated as a thermal average by means of the statistical operator
of, say, the canonical ensemble. Having in mind the scaling limits towards
conformal and massive field theories it is natural and useful to generalize both,
the notion of a local operator and the notion of the statistical operator. Instead
of local operators so-called quasi-local operators with tail were introduced in
\cite{BJMST06b,BJMST08a}. In these articles the XXZ chain, a model of locally
interacting spins, was considered on an infinite lattice. On such a lattice the
action of $S(k) = \frac 12 \sum_{j = - \infty}^k \s_j^z$, where $\s^z$ is a Pauli
matrix, makes sense. A quasi-local operator is an operator of the form
$q^{2\al S(0)}\mathcal{O}$, where $\mathcal{O}$ is local. Expectation values of
local operators (of spin zero) can be defined by means of a functional
\begin{equation} \label{Zkappa}
     Z^{\kappa, s} \Bigl\{q^{2\al S(0)}\mathcal{O} \Bigr\}
        = \frac{\Tr_{\mathrm{S}}\Tr_{\mathbf{M}}
      \Bigl\{Y_\mathbf{M}^{(-s)}T_{\mathrm{S},\mathbf{M}}\ q^{2\kappa S(\infty)} \
            \mathbf{b}^*_{\infty,s-1}\cdots \mathbf{b}^*_{\infty, 0}
        \bigl( q^{2 \al S(0)}\mathcal{O}\bigr)\Bigr\}}
           {\Tr _{\mathrm{S}}\Tr_{\mathbf{M}}\Bigl\{Y_\mathbf{M}^{(-s)}
             T_{\mathrm{S},\mathbf{M}}\ q^{2\kappa S(\infty)} \
         \mathbf{b}^*_{\infty,s-1}\cdots \mathbf{b}^*_{\infty, 0}
         \bigl( q^{2 \al S(0)}\bigr)\Bigr\}}
\end{equation}
generalizing the canonical ensemble average of statistical mechanics.
The `lattice screening operators' $\mathbf{b}^*_{\infty,j}$, which are
the coefficients in the expansion of the singular part of $\bb^*(\z)$ at $\z^2=0$,%
\footnote{For the case $s<0$ one can replace the operators $\bb^*_{\infty,j}$
by $\bc^*_{\infty,j}$ as in \cite{JMS11}.} increase the spin of $\mathcal{O}$
by $s$. This is compensated by the operator $Y_\mathbf{M}^{(-s)}$ at the boundary
which carries spin $- s$ and ensures that the ice-rule of the six-vertex model
is satisfied. As was discussed in \cite{BJMS10}, the functional (\ref{Zkappa})
does not depend on the concrete choice of $Y_\mathbf{M}^{(-s)}$. The spin is
needed as an additional parameter, necessary to exhaust the full space of
descendents in the conformal limit.

The `thermodynamic properties' of the average are determined by the monodromy
matrix $T_{\mathrm{S},\mathbf{M}}$ and the operator $q^{2\kappa S(\infty)}$.
Mathematically the monodromy matrix is defined by evaluating the universal
$R$-matrix of $U_q(\slth)$ on the tensor product of two evaluation representations
$\mathfrak{H}_{\mathrm{S}}$ and $\mathfrak{H}_{\mathbf{M}}$. Both of them realize
the space of states of a spin chain. With $\mathfrak{H}_{\mathrm{S}}$ we associate
an infinite chain and with $\mathfrak{H}_{\mathbf{M}}$ a finite chain of length $N$,
$$
     T_{\mathrm{S},\mathbf{M}}=\raisebox{.7cm}{$\curvearrowright $}
        \hskip -.75cm\prod\limits_{j=-\infty}^{\infty} T_{j,\mathbf{M}} \epc \qd
    T_{j,\mathbf{M}}\equiv T_{j,\mathbf{M}}(1) \epc \qd
    T_{j,\mathbf{M}}(\z)=\raisebox{.7cm}{$\curvearrowleft $} \hskip -.6cm
    \prod\limits_{\mathbf{m=1}}^{N} L_{j,\mathbf{m}}(\z/\x_m) \epp
$$
Here the $\x_j \in {\mathbb C}$ are inhomogeneity parameters, and $L$ is the
standard $L$-operator of the six vertex model
$$
L_{j,\mathbf{m}}(\zeta) =q^{-\frac 1 2\sigma ^3_j\sigma
^3_\mathbf{m}} -\z ^2 q^{\frac 1 2\sigma ^3_j\sigma ^3_\mathbf{m}}
-\z (q-q^{-1}) (\sigma ^+_j\sigma ^-_\mathbf{m}+\sigma ^-_j\sigma
^+_\mathbf{m})\,.
$$
Different physical realizations of the spin chain (e.g.\ finite temperature
or finite length) can be realized by appropriate choices of the parameters
$\x_j$ and $\k$ \cite{BoGo09}. These parameters are also at our disposal
for realizing scaling limits towards CFT \cite{BJMS10} or massive integrable
quantum field theories \cite{JMS11pp,JMS11b,JMS11}.

For the direction of infinite extension of the lattice one can change the
boundary conditions and, instead of taking the traces at the right hand side
of (\ref{Zkappa}), insert two one-dimensional projectors $|\kappa\rangle\langle\kappa|$
and $|\kappa+\al-s,s\rangle\langle\kappa+\al-s,s|$ at the boundary. Then
\begin{align}
&Z^{\kappa, s} \Bigl\{q^{2\al S(0)}\mathcal{O} \Bigr\}\rightarrow
\frac{\langle\kappa+\al-s,s|T_{\mathrm{S},\mathbf{M}}\ q^{2\kappa S}
\ \mathbf{b}^*_{\infty,s-1}\cdots \mathbf{b}^*_{\infty, 0} \bigl(
q^{2 \al S(0)}\mathcal{O}\bigr)|\kappa\rangle}
{\langle\kappa+\al-s,s|T_{\mathrm{S},\mathbf{M}}\ q^{2\kappa S} \
\mathbf{b}^*_{\infty,s-1}\cdots \mathbf{b}^*_{\infty, 0} \bigl( q^{2
\al S(0)}\bigr)|\kappa\rangle} \label{Zkappa1} \epc
\end{align}
where $|\kappa\rangle$ is the eigenvector of the transfer matrix
$T_\mathbf{M}(\z,\kappa)=\text{Tr}_j\bigl(T_{j,\mathbf{M}}q^{\kappa\sigma^3_j}\bigr)$
with maximal eigenvalue $T(\z,\kappa)$ in the spin-zero sector, and the
eigenvector $|\kappa+\al-s,s\rangle$ corresponds to the maximal eigenvalue
$T(\z,\kappa+\al-s,s)$ of the transfer matrix $T_\mathbf{M}(\z,\kappa+\al-s)$
in the sector with spin $s$.\footnote{For $s \ne 0$, depending on the choice
of the inhomogenieties, the eigenvalues may be generally degenerate. This is,
however, irrelevant for our discussion below.}
In \cite{BJMS10} we argued that the combination
\begin{equation} \label{kappa'}
     \kappa'=\kappa +\al+2\textstyle {\frac {1-\nu}\nu}s
\end{equation}
becomes an independent parameter in the CFT scaling limit.

\subsection{Functions $\rho, \om$ and basic thermodynamic functions.}
In \cite{BJMST08a} the `fermionic basis' described in the introduction was
constructed. The creation operators $\tb^*$, $\bb^*$ and $\cb^*$ generate
the space of quasi-local operators by acting on the pure tail $q^{2 \a S(0)}$
which plays the role of the Fock vacuum for these operators. Note that the
creation operators are very special. Their most important property is their
compatibility with the above defined functional $Z^{\k, s}$ revealing itself
in the following formulae proved in \cite{JMS08}
\begin{align}
&Z^{\kappa,s}\bigl\{\tb^*(\z)(X)\bigr\}
=2\rho(\z|\kappa,\kappa+\al,s)Z^{\kappa,s}\{X\}\,,\label{JMS}\\
&Z^{\kappa,s}\bigl\{\bb^*(\z)(X)\bigr\} =\frac 1{2\pi i}\oint\limits
_{\Gamma} \omega (\z,\xi|\kappa,\al,s)
Z^{\kappa,s}\bigl\{\cb(\xi)(X)\bigr\}
\frac{d\xi^2}{\xi^2}\,,\nn\\
&Z^{\kappa,s}\bigl\{\cb^*(\z)(X)\bigr\} =-\frac 1 {2\pi
i}\oint\limits_{\Gamma} \omega (\xi,\z|\kappa,\al,s)
Z^{\kappa,s}\bigl\{\bb(\xi)(X)\bigr\} \frac{d\xi^2}{\xi^2}\epp
\nn
\end{align}
Here the contour $\Gamma$  encircles all the singularities of the integrand
except $\xi^2=\z^2$. The functions $\r$ and $\om$ appearing in this theorem
are in the center of interest of this work. We shall provide a precise definition
below. They are fundamental for the description of all static correlation
functions of the XXZ chain and its various scaling limits, since Wick's theorem
combined with (\ref{JMS}) implies the determinant formula
\begin{multline}
     Z^{\kappa,s}\bigl\{\tb^*(\z^0_1)\cdots \tb^*(\z^0_p)
     \bb^*(\z^+_1)\cdots \bb^*(\z^+_r) \cb^*(\z^-_r)\cdots
     \cb^*(\z^-_1)\bigl(q^{2\al  S(0)}\bigr)\bigr\} \label{det} \\
     =\prod\limits _{i=1}^p 2\rho (\z _i^{0}|\kappa,\kappa+\al,s)\times
      \det \left(\omega(\z^+_i,\z ^-_j|\kappa,\al,s) \right)_{i,j=1,\cdots, r} \epp
\end{multline}
From this formula one can obtain any correlation function by Taylor expanding
both sides and comparing coefficients, because the operators appearing on the
left hand side generate a basis of the space of quasi-local operators as was
proved in \cite{BJMS09a}.

The function $\rho$ in (\ref{JMS}) is the ratio of two eigenvalues of the transfer
matrix
\begin{equation} \label{rho}
     \rho(\z|\kappa,\kappa+\al,s)=\frac{T(\z,\kappa+\al-s,s)}{T(\z,\kappa)} \epp
\end{equation}
In the following we shall replace it with
\begin{equation} \label{rho1}
     \rho(\z|\kappa,\kappa') = \frac{T(\z,\kappa')}{T(\z,\kappa)}
\end{equation}
and treat $\k'$ as real parameter independent of $\a$ (which reappears below in
the definition of $\om$). As was explained in Section 4 of \cite{BJMS10} this
replacement is possible and necessary in the CFT scaling limit.


For the function $\om$ the following representation was originally
obtained\footnote{In \cite{BoGo09} we used a slightly different
notation. In particular, up to some multiplier, the first two terms in
brackets on the right hand side of (\ref{om}) were denoted $\Psi$.}
in \cite{BoGo09} and then used in \cite{BJMS10},
\begin{equation} \label{om}
     \frac14\omega(\z,\z'|\kappa,\kappa';\alpha) =
        \Bigl( f_\mathrm{left}\star f_\mathrm{right}
           + f_\mathrm{left}\star R \star f_\mathrm{right}\Bigr)(\z,\z')
           + \omega_0(\z,\z'|\al) \epp
\end{equation}
In the CFT scaling limit the function $\omega(\z,\z'|\kappa,\kappa';\alpha)$ is
identical to $\omega(\z,\z'|\kappa,\al,s)$ in (\ref{JMS}) if (\ref{kappa'}) is
fulfilled \cite{BJMS10}.

The definition of the functions entering into (\ref{om}) is slightly involved.
But they either derive from the elementary function
\begin{equation} \label{psi}
     \psi (\z,\al)=\z^{\al}\frac{\z^2 +1}{2(\z^2-1)}
\end{equation}
or from the functions appearing in the $TQ$-relation. For their definition
we shall also need the two difference operators
\begin{equation} \label{deltas}
     \Delta _\z f(\z) = f(\z q)-f(\z q^{-1}) \epc \qd
     \delta ^-_\z f(\z) = f(\z q) -\rho(\z|\kappa,\kappa') f(\z) \epp
\end{equation}
Then
\begin{equation} \label{flfr}
     f_\mathrm{left}(\z,\z',\alpha) =
        \frac{1}{2\pi i} \delta_\z^-\psi(\z/\z',\alpha) \epc \quad
     f_\mathrm{right}(\z,\z',\alpha) = \delta_{\z'}^-\psi(\z/\z',\alpha)
\end{equation}
and
\begin{equation}
     \omega _0(\z,\z'|\al) =\delta_\z^-\delta_{\z'}^-\Delta ^{-1}_\z
       \psi(\z/\z',\alpha) \epc
\end{equation}
where $\Delta^{-1}\psi$ is defined as a principal value integral
\begin{equation} \label{invdeltapsi}
     \Delta ^{-1}_\z \psi(\z,\alpha) = - PV \int\limits _0^{\infty}
        \frac{\rd \eta ^2}{2\pi i\eta ^2} \:
        \frac{\psi(\eta,\al)}{2\nu\bigl(1+(\z/\eta)^{\frac 1 \nu}\bigr)} \epc \qd
    \z^2>0 \epc \quad -\frac 1 \nu<\mathop{{\rm Re}}\al<0 \epp
\end{equation}

The function $R$ is the resolvent of a linear integral operator. For its
definition we introduce the kernels
\begin{equation} \label{defK}
     K_{\al}(\z) = \frac 1{2\pi i}\Delta _{\z}\psi (\z,\al) \epc \quad
     K(\z)=K_0(\z) \epp
\end{equation}
Then $R$ is defined by the integral equation
\begin{align} \label{eqR}
     R - R \star K_\alpha = K_\alpha \epp
\end{align}
Here and in (\ref{om}) we used the notation
$$
     (f\star g)(\z,\z')= \int \limits_\gamma \rd m(\eta) \: f(\z,\eta)g(\eta,\z')
$$
for the convolution. The `measure' $\rd m$ is given as
\begin{align} \label{measure}
     \rd m (\eta ) = \frac {\rd \eta ^2}{ \eta ^2\rho (\eta|\kappa, \kappa')
                     \bigl(1+\mathfrak{a}(\eta,\kappa)\bigr)} \epc
\end{align}
and the contour $\gamma$ is described below.

The auxiliary function $\mathfrak{a}$ is defined by
\begin{equation} \label{defa}
     \mathfrak{a}(\z,\kappa)=\frac{d(\z)Q(q\z,\kappa)}{a(\z)Q(q^{-1}\z,\kappa)}
\end{equation}
where $Q(\z,\kappa)$ is the eigenvalue of the $Q$-operator (for the definition
of the $Q$-operator see Section 3 of \cite{BJMS10}). It satisfies Baxter's TQ-equation
\begin{align}
T(\z,\kappa)Q(\z,\kappa)=d(\z)Q(q\z,\kappa)+a(\z)Q(q^{-1}\z,\kappa)\label{TQ}
\end{align}
where $q=\exp{(\pi\nu i)}$ and
\begin{align} \label{ad}
     a(\z) = \prod_{j=1}^N(1-q\z^2/\xi_j^2) \epc \quad
     d(\z)=\prod_{j=1}^N(1-q^{-1}\z^2/\xi_j^2) \epp
\end{align}
For our convenience we assume $N$ to be even. The function $\z^\k Q(\z,\kappa)$
is a polynomial and, hence, is determined by its zeros $\z_j$ (the Bethe roots)
satisfying the Bethe ansatz equation
$$
     \mathfrak{a}(\z_j,\kappa)=-1, \quad j=1,\cdots,N/2 \epp
$$

Instead of (\ref{defa}) one can use a non-linear integral equation to characterize
the auxiliary function $\fa$ \cite{Kluemper92,Kluemper93}. Its precise form depends
on the position of the inhomogeneity parameters \cite{BoGo09}. The non-linear
integral equation is particularly useful for numerical calculations, for the
calculation of the large $N$ asymptotics and for performing the limit $N
\rightarrow \infty$ in the temperature case
\begin{align}
     \xi_{2j-1} = \exp{(\pi\nu i/2-\beta/N)} \epc \quad
     \xi_{2j} =\exp{(-\pi \nu i/2+\beta/N)} \epc \quad j=1, \dots, N/2 \epp
\end{align}
In \cite{BJMS10} we considered the conformal limit. In the present
paper we deal with the temperature case
with the inverse temperature $\beta=T^{-1}$. Then, after taking the limit
$N\to\infty$, the non-linear equation for the auxiliary function becomes
\begin{equation} \label{non-linearT}
     \log \bigl(\mathfrak{a}(\z,\kappa)\bigr)
        = -2i\pi \nu\kappa - i\sin{(\pi \nu)}\beta e(\z)\; -\;
      \int\limits_\gamma \frac {\rd\xi^2}{\xi ^2} \:
      K(\z/\xi)\log \bigl(1+ \mathfrak{a}(\xi,\kappa)\bigr)
\end{equation}
with the `bare energy' $e(\z)=e(\exp{\la})=\coth{(\la)}-\coth{(\la+\pi  \nu i)}$.
The contour $\gamma$ in the complex $\xi^2$ plane encompasses the essential
singularity at $\xi^2=1$, but the poles of the kernel at $\x = \z q^{\pm 1}$
are outside. In terms of the variable $\mu=\log{\xi}$ it is depicted
in \cite{BoGo09} with the opposite integration direction. In the following
we assume $\k$ and $\n$ to be real. This corresponds to the unphysical case of
a purely imaginary magnetic field, but is convenient for our purpose of
studying the function $\om$ by means of the high-temperature expansion.

\subsection{Functions $G$ and $\overline G$}
Originally the correlation functions in the temperature case for $\a = 0$
were represented by means of a function $G$ introduced in \cite{GKS04a,GKS05}. The
generalization to $\al\ne 0$ was achieved in \cite{BoGo09}. Using the above
notation, we can define a generalized $G$-function and its `dual' $\overline G$
through the linear integral equations
\begin{align}
     & G = f_\mathrm{right} +  K_{\al} \star G \epc \label{eqG} \\
     & \overline G = f_\mathrm{left}+ \overline G \star K_{\al} \epp \label{eqbarG}
\end{align}
A formal solution utilizing the resolvent is
\begin{align}
     & G(\z,\z'|\kappa,\kappa';\alpha) = \Bigl(f_\mathrm{right}+  R \star
       f_\mathrm{right}\Bigr)(\z,\z') \epc \label{G}\\
     & \overline G(\z,\z'|\kappa,\kappa';\alpha) = \Bigl(f_\mathrm{left}+
       f_\mathrm{left}\star R\Bigr)(\z,\z') \epp \label{barG}
\end{align}
Up to some factor we reproduce the functions $G$ and
$$
     \Psi(\z,\z'|\kappa,\kappa';\alpha) = \Bigl(f_\mathrm{left}\star G\Bigr)(\z,\z')
        =\Bigl(\overline G\star f_\mathrm{right}\Bigr)(\z,\z') \epp
$$
from \cite{BoGo09} for $\kappa' = \a + \k$. In the general case $\Psi$ is equal to
the terms in brackets on the right hand side of (\ref{om}).

The important `normalization condition' for the function $\om$, originally
suggested in \cite{JMS08}, includes the asymptotics for large spectral
parameter
\begin{equation}
     \lim_{\z\to\infty} \z^{-\a} \om(\z,\z'|\kappa,\kappa + \a;\al) = 0 \epp
\end{equation}
In \cite{BoGo10} we showed this directly from the definition of the previous
section using the properties of the linear integral equation (\ref{eqG}).
The key identity in our proof was
\begin{equation} \label{identG}
    \lim_{\z\to\infty}2 (\z/\z')^{- \a} G(\z,\z')
       = q^{-\al}-\rho(\z')
         -\int_{\gamma} \frac{\rd m(\eta)}{2 \p i} \:
	 G(\eta,\z') \D _\h (\eta/\z')^{-\a} = 0 \epc
\end{equation}
valid for $\k' = \k + \a$ in the measure.

\subsection{Functions $\sigma$ and $\Phi$}
\label{subsec:sigmaphi}
In order to prove (\ref{identG}) we introduced a generalized `dressed charge'
$\sigma$ and a function $\Phi$ in \cite{BoGo10}. Here we will use slightly
modified and generalized definitions of these functions adapted to the notation
in \cite{BJMS10}.
Namely, we define a function $\sigma$ which satisfies the following
linear integral equation
\begin{equation} \label{eqsigma}
     \sigma(\z|\kappa,\kappa';\al) = \D_\z \z^{-\al}
        + \int_{\gamma}\rd m(\eta) \: \sigma(\eta|\kappa,\kappa';\al)
                   K_{\al}(\eta/\z) \epp
\end{equation}
The so-called `dressed function trick' implies that
$$
     \int_{\gamma}\rd m(\eta) \: G(\eta,\z'|\kappa,\kappa';\al) \D_\h \eta^{-\al} =
        \int_{\gamma}\rd m(\eta) \: \sigma(\eta|\kappa,\kappa';\al)
	f_\mathrm{right}(\eta,\z')
    \epp
$$
Then (\ref{identG}) is reduced to an identity for $\sigma$ with $\k' = \k + \a$
which can be proved with the help of the properties of the function
\begin{equation} \label{Phi}
     \Phi(\z|\kappa,\kappa') = \frac{Q(\z,\kappa')}{Q(\z,\kappa)} \epp
\end{equation}

The key observation here is that two TQ-equations (\ref{TQ}) with twist parameters
$\kappa$ or $\kappa'$, respectively, can be combined into the following relation
for $\PH$,
\begin{equation} \label{relPhi}
     \frac{1}{\rho(\eta|\kappa,\kappa')(1+\mathfrak{a}(\eta,\kappa))} =
        - \frac{\Phi(\eta|\kappa,\kappa')}{\D_\h \Phi(\eta|\kappa,\kappa')}
        + \frac{\Phi(q\eta|\kappa,\kappa')}
           {\rho(\eta|\kappa,\kappa') \D_\h \Phi(\eta|\kappa,\kappa')}
\end{equation}
Introducing an operator
\begin{equation} \label{H}
     H_{\eta}
     = \frac{1}{1+\bar{\mathfrak{a}}(\eta,\kappa)}\; d^+_{\eta}
       +\frac{1}{1+\mathfrak{a}(\eta,\kappa)} \;
        d^-_{\eta}-\rho(\eta|\kappa,\kappa') \epc
\end{equation}
where $d^{\pm}_{\eta}f(\eta)=f(q^{\pm 1}\eta)$ and
$\bar{\mathfrak{a}}(\eta,\kappa)= 1/\mathfrak{a}(\eta,\kappa)$, we can rewrite this
more compactly as
\begin{align} \label{relPhi1}
     H_{\eta}\Phi(\eta|\kappa,\kappa') = 0 \epp
\end{align}
This is the operator $H$ which we mentioned already in the introduction and
which will be useful for the characterization of our main function $\om$ as
well.

For $\k' = \k + \a$ the solution of the integral equation (\ref{eqsigma}) for
$\sigma$ can be expressed in terms of the function $\Phi$:
\begin{align} \label{solsigma}
     \sigma(\z) = \Delta_{\z} \Phi(\z)/ \Phi_0
\end{align}
with some known normalization constant $\Phi_0$ \cite{BoGo10}. This was shown
in \cite{BoGo10} for the inhomogeneous model with finite $N$. Here we shall
repeat the argument for the temperature case.

For the proof we need to know the location of the singularities of
the involved functions. If $N$ is finite the function $\Phi(\z)$ has
poles at the Bethe roots corresponding to the zeros of the
denominator $Q(\z,\kappa)$. It follows from our definition of the
$Q$-operator described in Section 3 of \cite{BJMS10} that
\begin{equation} \label{asQ}
     Q(\z,\kappa) = \z^{-\kappa+s}A(\z,\kappa) \epc
\end{equation}
where $A(\z,\kappa)$ is a rational function in $\z^2$.
As explained above we consider the sector $s=0$ here. Thus,
\begin{equation} \label{asPhi}
     \Phi(\z|\kappa,\kappa')/\Phi_0=
        \z^{-\kappa'+\kappa}\phi(\la|\kappa,\kappa') \epc
\end{equation}
where $\la=\log{\z}$ by definition and where $\phi(\la|\kappa,\kappa')$ has
constant asymptotics for $\Re \la \rightarrow \pm \infty$.

It is well known that in the temperature case all Bethe roots `condense'
to the point $\z=1$ which becomes an essential singularity of
$\Phi(\z|\kappa,\kappa')$.
We conclude that $\phi(\la|\kappa,\kappa')$ has an expansion of the form
\begin{equation} \label{expanphi}
     \phi(\la|\kappa,\kappa')=1+\sum\limits_{j\ge 0}c_j\cth^{(j)}(\la) \epc
\end{equation}
where $\cth^{(j)}(\la):=(\6/\6{\la})^j\cth(\la)$ and where the dependence on the
inverse temperature $\be$ is in the coefficients $c_j$ which also depend on
$q,\kappa,\kappa'$. Note that all partial sums in \eqref{expanphi} are still
meromorphic in the original variable $\z^2$. The coefficients $c_j$ have the
high-temperature expansions
\begin{equation} \label{tempc}
     c_j = \sum\limits_{k>j}\be^kc_{k|j}
\end{equation}
with respect to $\be$. The most convenient way to calculate the coefficients
$c_{k|j}$ is to consider the functional equation
\begin{equation} \label{Phiauxfun}
     \frac{\Phi(q\z|\kappa,\kappa')}{\Phi(q^{-1}\z|\kappa,\kappa')}
        = \frac{\mathfrak{a}(\z,\kappa')}{\mathfrak{a}(\z,\kappa)} \epc
\end{equation}
where the high-temperature expansion on the right hand side follows directly
form the integral equation (\ref{non-linearT}). In Appendix \ref{app:htphi}
we will explain how to calculate the lowest coefficients explicitly.

For $\kappa'=\kappa+\al$ we have $ \Phi(\eta| \k, \k + \a)/\Phi_0 =
\eta^{-\al} \phi(\eta| \k, \k + \a)$, and the factor $\eta^{-\al}$ cancels
the corresponding factor under the integral in (\ref{eqsigma}) stemming from
the kernel $K_{\al}(\eta/\z)$. Then, if we substitute (\ref{relPhi}) into the
integral and take into account that both, the function $\sigma(\z)$ and the second
term in the right hand side of (\ref{relPhi}), do not have singularities inside
the contour, we conclude that this second term does not contribute to the
integral. Hence, it can be dropped and the integral can be calculated by
deforming the contour as it was explained in \cite{BoGo10}.

\section{Solving the generalized dressed charge equation by means of the
high-temperature expansion}
We are now going to generalize the above ideas step by step. We will show
that all the linear integral equations considered above, like equation
(\ref{eqsigma}) 
for the function $\sigma$, equations (\ref{eqG}), (\ref{eqbarG}) for the
$G$-functions or equation (\ref{eqR}) for the resolvent, can be solved in terms
of a single function of one or two spectral parameters with essential
singularities if one of the spectral parameters equals 1 and a certain
behavior at 0 and $\infty$, as in the above example of the function
$\sigma(\z)$ at $\kappa'=\kappa+\al$. We shall see in the next section that
one such function $\Phi(\z,\z')$ with essential singularities at $\z=\z'=1$
will be all we need. The full information about the solutions of the above
mentioned linear integral equations is contained in this single function.

Before we proceed to the general case let us describe how our method works
for equation (\ref{eqsigma}) with independent parameters $\kappa,\kappa'$ and
$\al$. We have seen above that $\sigma$ given by (\ref{solsigma}) with
$\Phi(\z|\kappa,\kappa')$ defined by (\ref{Phi}) solves equation (\ref{eqsigma})
only if $\kappa'=\kappa+\al$. The problem with the general case is that the
behavior at $\z=0,\infty$ does not match the corresponding behavior of the
kernel $K_{\al}$ in the integral of (\ref{eqsigma}). Hence, we need a generalized
function $\Phi(\z|\kappa,\kappa';\al)$.

In order for a formula similar to (\ref{solsigma}) to work, we first of all
need a relation similar to (\ref{relPhi}):
\begin{equation} \label{genrelPhi}
     \frac{1}{\rho(\eta|\kappa,\kappa')(1+\mathfrak{a}(\eta,\kappa))}
        = -\frac{\Phi(\eta|\kappa,\kappa';\al)}{\D_\h \Phi(\eta|\kappa,\kappa';\al)}
      + r(\eta|\kappa,\kappa';\al) \epc
\end{equation}
where the remainder function $r$ is not defined yet. The only property of
this function is that it does not have singularities inside the
contour $\gamma$. We will call such functions `regular'. We also
need a generalization of the asymptotic property and the expansion
(\ref{asPhi}), (\ref{expanphi}),
\begin{equation} \label{genasPhi}
     \Phi(\z|\kappa,\kappa';\al)= \z^{-\al}\phi(\la|\kappa,\kappa';\al) \epc
        \quad \la=\log{\z}
\end{equation}
with
\begin{equation} \label{expanphi'}
     \phi(\la|\kappa,\kappa';\al)=1+\sum\limits_{j\ge 0}c'_j\cth^{(j)}(\la) \epc
\end{equation}
where the coefficients $c'_j$ now depend on $q,\kappa,\kappa', \al$
and on the inverse temperature $\be$. The high-temperature expansion
is of the same form as (\ref{tempc}),
\begin{equation} \label{tempc'}
     c'_j = \sum\limits_{k>j}\be^kc'_{k|j} \epp
\end{equation}

The relations (\ref{relPhi}) and (\ref{genrelPhi}) can be combined into
\begin{equation} \label{genrelPhi1}
     \frac{\Phi(\eta|\kappa,\kappa')}{\D_\h \Phi(\eta|\kappa,\kappa')}
        - \frac{\Phi(\eta|\kappa,\kappa';\al)}{\D_\h \Phi(\eta|\kappa,\kappa';\al)}
    = \tilde{r}(\eta|\kappa,\kappa';\al)
\end{equation}
with a regular remainder $\tilde r$. Our first claim is that this equation has
a unique solution in terms of the expansion (\ref{expanphi'}). Our second claim
is that, due to similar arguments as at the end of the previous Section, the
solution to the equation (\ref{eqsigma}) can be expressed as
\begin{equation} \label{soleqsigma}
     \sigma(\z|\kappa,\kappa';\al) = \Delta_{\z}\Phi(\z|\kappa,\kappa';\al) \epp
\end{equation}

We can interpret (\ref{genrelPhi1}) as a cancellation condition for the singular
part at $\la=0$ of the functions at the left hand side. Inserting (\ref{asPhi}) and
(\ref{genasPhi}) we find that the following  expression is regular (which we denote
by `$= \text{reg}$'),
\begin{multline} \label{cond}
     \phi(\la|\kappa,\kappa')\bigl(q^{-\a}\phi(\la+\pi \nu i|\kappa,\kappa';\al)
        -q^{\a}\phi(\la-\pi \nu i|\kappa,\kappa';\al)\bigr) \\
    - \phi(\la|\kappa,\kappa';\al)
      \bigl(q^{-\kappa'+\kappa}\phi(\la+\pi \nu i|\kappa,\kappa')
           -q^{\kappa'-\kappa}\phi(\la-\pi \nu i|\kappa,\kappa')\bigr)
        = \text{reg} \epp
\end{multline}
We substitute the expansions (\ref{expanphi}), (\ref{expanphi'}) into here and
use the elementary formula
$$
     \cth(\la)\cth(\la')=1+\cth(\la)\cth(\la'-\la)-\cth(\la')\cth(\la'-\la)
$$
to extract the singular part with respect to $\la$ for both terms on the left
hand side of (\ref{cond}). Comparing the coefficients in front of corresponding
terms $\cth^{(j)}(\la)$ we obtain the following equation for $c_j, c'_j$,
\begin{multline} \label{eqcc'}
   (q^{\a}-q^{-\a})c_j-(q^{\kappa'-\kappa}-q^{-\kappa'+\kappa}) c'_j= \\
      \sum\limits_{j\; ',j\; ''\ge 0}c_{j'} c'_{j''}
          \biggl((-1)^{j''}\binom{j'}{j}\gamma_{j'+j''-j}(\al)
         -(-1)^{j'}\binom{j''}{j}\gamma_{j'+j''-j}(\kappa'-\kappa)\biggr) \epc
\end{multline}
where
\begin{equation} \label{gamma}
     \gamma_j(\al):=(q^{\al}+(-1)^j q^{-\al})\cth^{(j)}(\pi \nu i)
\end{equation}
and where the binomial coefficients are defined in such a way that $\binom{a}{b}=0$
for $a<b$. If $\kappa'=\kappa+\al$ the unique solution is $c'_j=c_j$, since both
sides of (\ref{eqcc'}) vanish identically.

Further inserting the high-temperature expansions (\ref{tempc}), (\ref{tempc'})
we obtain
\begin{multline} \label{tempeqcc'}
     (q^{\a}-q^{-\a}) c_{k|j} - (q^{\kappa'-\kappa} - q^{-\kappa'+\kappa}) c'_{k|j} =
        \\
     \sum\limits_{k'=1}^{k-1}\;\sum\limits_{j\;'=0}^{k'-1}\;
     \sum\limits_{j\;''=0}^{k-k'-1} c_{k'|j'}c'_{k-k'|j''}
     \biggl[(-1)^{j''}\binom{j'}{j}\gamma_{j'+j''-j}
     (\al)-(-1)^{j'}\binom{j''}{j}\gamma_{j'+j''-j}(\kappa'-\kappa)\biggr]
\end{multline}
for $0\le j\le k-1$.

If the coefficients $c_j$ are known up to a certain order $k$ of the
high-temperature expansion (\ref{tempc}) 
we can get the $c'_j$ from the above relation (\ref{tempeqcc'}) up to the same
order, because all coefficients on the right hand side are of lower order.
Thus, one can iteratively solve equation (\ref{tempeqcc'}) with respect to
the coefficients $c'_j$, once the $c_j$ are known. The solution obtained in this
way is obviously unique. In Appendix \ref{app:htgenphi} we show the lowest order
coefficients explicitly.

\begin{rem*}
Instead of the relation (\ref{genrelPhi1}) we could consider, for example,
\begin{equation} \label{alminusal}
     \frac{\Phi(\eta|\kappa,\kappa';\al)}{\D_\h \Phi(\eta|\kappa,\kappa';\a)}
        - \frac{\Phi(\eta|\kappa,\kappa';-\al)}{\D_\h \Phi(\eta|\kappa,\kappa';-\al)}
    = \text{reg} \epp
\end{equation}
Then, repeating the above arguments, we come to the following compatibility condition
\begin{equation} \label{eqcbarc}
     (q^{\a}-q^{-\a})(c'_j+\bar{ c}'_j)
        = -\sum\limits_{j\; ',j\; ''\ge 0}(-1)^{j'}c'_{j'} \bar{c}'_{j''}
       \biggl((-1)^j\binom{j'}{j}-\binom{j''}{j}\biggr)\gamma_{j'+j''-j}(\a) \epc
\end{equation}
where the coefficients $\bar c'_j$ appear in
\begin{equation} \label{expanphiminusal}
     \phi(\la|\kappa,\kappa';-\al) = 1+\sum\limits_{j\ge 0}\bar{c}'_j\cth^{(j)}(\la)
\end{equation}
and have the high-temperature expansion
\begin{equation} \label{tempbarc'}
     \bar c'_j = \sum\limits_{k>j}\be^k \bar c'_{k|j} \epp
\end{equation}
\end{rem*}

\section{Resolvent and master function}
We shall show in this section that the arguments applied above in order to
obtain a solution $\s$ of the dressed charge equation (\ref{eqsigma}) in terms
of a function $\Phi$, which has a simple high-temperature expansion, can be
generalized to the linear integral equation (\ref{eqR}) for the resolvent $R$.
To this end we introduce another function $\Phi(\z,\z'|\kappa,\kappa';\a)$ now
depending on two spectral parameters $\z,\z'$. As we shall see, this function
contains the whole information which is necessary for the description of all
functions defined by means of linear integral equations that appeared before.
For this reason it may be called a `master function'.

Generalizing the ideas of the previous section we define it as follows:
\begin{align} \label{masterPhi}
     & \Phi(\z,\z'|\kappa,\kappa';\a) =
        \Phi'(\z,\z'|\kappa,\kappa';\a)+\Delta_{\z}^{-1}\psi(\z/\z',\a) \epc \\
     & \Phi'(\z,\z'|\kappa,\kappa';\a) =
        \frac{1}{2}\bigl(\z/\z'\bigr)^{\a}\phi(\la,\la'|\kappa,\kappa';\a) \epc \quad
        \la=\log{(\z)} \epc \quad \la'=\log{(\z')} \epc \label{asmasterPhi}
\end{align}
where
\begin{align}
\phi(\la,\la'|\kappa,\kappa';\a)=\sum\limits_{j,j'\ge 0}
c_{j,j'}\cth^{(j)}(\la)\cth^{(j')}(\la') \label{phi2expan}
\end{align}
and where $\Delta^{-1}_\z \psi(\z, \a)$ was defined in (\ref{invdeltapsi}).
We complete the definition by demanding that $\Phi(\z,\z'|\kappa,\kappa';\a)$
satisfies an equation similar to (\ref{genrelPhi}),
\begin{align}
     \frac{1}{\rho(\eta|\kappa,\kappa')(1+\mathfrak{a}(\eta,\kappa))} =
        & - \frac{\Phi'(\eta,\z'|\kappa,\kappa';\al)}
             {\D_\h \Phi(\eta,\z'|\kappa,\kappa';\al)} +
          r(\eta,\z'|\kappa,\kappa';\al) \label{masterrelPhi1}\\
      = & - \frac{\Phi'(\z,\eta|\kappa,\kappa';\al)}
                 {\D_\h \Phi(\z,\eta|\kappa,\kappa';\al)} +
          r'(\z,\eta,|\kappa,\kappa';\al) \epc \label{masterrelPhi2}
\end{align}
where the remainders $r,r'$ are regular functions of $\eta$.

We shall see below that (\ref{asmasterPhi}) and (\ref{masterrelPhi1}),
(\ref{masterrelPhi2}) determine the coefficients $c_{j, j'}$ in the
high-temperature expansion (\ref{phi2expan}). The relation of our master
function with the resolvent is explained in the following
\begin{lem} \label{lem:R}
The resolvent $R$ defined by the linear integral equation (\ref{eqR})
can be represented as
\begin{equation} \label{solR}
     R(\z,\z'|\kappa,\kappa';\al) =
        - \frac{1}{2\pi i}\Delta_{\z}\Delta_{\z'}\Phi(\z,\z'|\kappa,\kappa';\al) \epp
\end{equation}
\end{lem}
A proof of this Lemma is provided in Appendix \ref{app:proofresolvent}.

The function $\Phi$ of one spectral parameter introduced in the previous Section
is recovered in the limits of large $\z$, $\z'$,
\begin{align}
     & \lim_{\z\to\infty}\z^{-\a}R(\z,\z'|\kappa,\kappa';\al)
        = - \frac{1}{4\pi i} \sigma(\z'|\kappa,\kappa';\al)
    = - \frac{1}{4\pi i}\Delta_{\z'}\Phi(\z'|\kappa,\kappa';\al) \epc
       \label{limits} \\
     & \lim_{\z'\to\infty}{\z'}^{\a}R(\z,\z'|\kappa,\kappa';\al)
        = - \frac{1}{4\pi i} \sigma(\z|\kappa,\kappa';-\al)
    = - \frac{1}{4\pi i}\Delta_{\z}\Phi(\z|\kappa,\kappa';-\al) \epc \notag
\end{align}
or
\begin{align} \label{limitPhi}
     & \lim_{\z\to\infty} \z^{-\a} \Phi(\z,\z'|\kappa,\kappa';\al)
        = \frac{\Phi(\z'|\kappa,\kappa';\al)}{2(q^\a - q^{- \a})} \epc \\
     & \lim_{\z'\to\infty} {\z'}^{\a} \Phi(\z,\z'|\kappa,\kappa';\al)
        = \frac{\Phi(\z|\kappa,\kappa';-\al)}{2(q^{- \a} - q^\a)} \epp \notag
\end{align}

Let us now come back to our claim that the high-temperature expansion
(\ref{phi2expan}) is determined by (\ref{masterrelPhi1}), (\ref{masterrelPhi2})
and by the asymptotic condition (\ref{asmasterPhi}). Combining (\ref{masterrelPhi1}),
(\ref{masterrelPhi2}) with (\ref{genrelPhi}) we obtain
\begin{align}
     & \frac{\Phi(\eta|\kappa,\kappa';-\al)}{\D_\h \Phi(\eta|\kappa,\kappa';-\al)} -
       \frac{\Phi'(\eta,\z'|\kappa,\kappa';\al)}
            {\D_\h \Phi(\eta,\z'|\kappa,\kappa';\al)} =
       \tilde{r}(\eta,\z'|\kappa,\kappa';\al) \epc \label{masterrelPhi1a} \\
     & \frac{\Phi(\eta|\kappa,\kappa';\al)}{\D_\h \Phi(\eta|\kappa,\kappa';\al)} -
       \frac{\Phi'(\z,\eta|\kappa,\kappa';\al)}
            {\D_\h \Phi(\z,\eta|\kappa,\kappa';\al)} =
       \tilde{r'}(\z,\eta|\kappa,\kappa';\al) \epc \label{masterrelPhi2a}
\end{align}
where the remainders $\tilde{r}$, $\tilde{r}'$ are also regular with respect to the
variable $\eta$.

First of all we shall see that (\ref{masterrelPhi1a}) and (\ref{masterrelPhi2a})
are equivalent to each other. We substitute (\ref{masterPhi}) into
(\ref{masterrelPhi2a}) and obtain the regularity condition
\begin{multline}
     \phi(\mu|\kappa,\kappa';\al)
        \biggl(q^{-\a}\phi(\la,\mu+\pi\nu i|\kappa,\kappa';\al)
          -q^{\a}\phi(\la,\mu-\pi\nu i|\kappa,\kappa';\al)-\cth(\la-\mu)\biggr)
          \label{phi2phi} \\
        - \phi(\la,\mu|\kappa,\kappa';\al)
      \biggl(q^{-\a}\phi(\mu+\pi\nu i|\kappa,\kappa';\al)
           -q^{\a}\phi(\mu-\pi\nu i|\kappa,\kappa';\al)\biggr) = \text{reg} \epc
\end{multline}
where $\mu=\log{(\eta)}$ and where the variable $\z$ corresponding to $\la=\log{(\z)}$
must be outside the integration contour $\gamma$.

If we repeat the procedure described in the previous Section, namely, if we
substitute the expansions (\ref{expanphi'}), (\ref{phi2expan}) into (\ref{phi2phi})
and into the equation obtained from (\ref{phi2phi}) by replacing $\a\to-\a$ and then
take the singular part, we obtain two equations for the expansion coefficients
$c_{j,j'}$
\begin{align}
     & \sum\limits_{j\;'\ge 0}c_{j,j'} U_{j',l}
        = \binom{j+l}{l}c'_{j+l} \epc \label{cU} \\
     & \sum\limits_{j\;'\ge 0}\bar U_{j,j'} (-1)^{j' + l} c_{j',l}
        = - (-1)^{j+l}\binom{j+l}{l}{\bar c}'_{j+l} \epc \label{cbarU}
\end{align}
where
\begin{align}
     & U_{j,l} = (q^{\a}-q^{-\a})\delta_{j,l}
        +\sum\limits_{j\;'\ge 0}c'_{j'}\gamma_{j,j'|l} \epc \label{U} \\
     & \bar U_{j,l} = (q^{\a}-q^{-\a})\delta_{j,l}
        -\sum\limits_{j\;'\ge 0}(-1)^{j'}{\bar c}'_{j'}\gamma_{l,j'|j} \label{barU}
\end{align}
and
\begin{align}
     \gamma_{j,j'|l} =
        \biggl((-1)^j\binom{j'}{l}-(-1)^{j'}\binom{j}{l}\biggr)\gamma_{j+j'-l}(\al)
\label{gamma1}
\end{align}
with $\gamma_j$ defined in (\ref{gamma}).

The compatibility condition for the equations (\ref{cU}), (\ref{cbarU}) is
$$
     \sum\limits_{j\;'\ge 0}(-1)^j\binom{j+j'}{j}{\bar c}'_{j+j'}U_{j'',l}
        = -\sum\limits_{j\;'\ge 0} \bar U_{j,j'}(-1)^{j'}\binom{j'+l}{l} c'_{j'+l} \epp
$$
Substituting (\ref{U}) and (\ref{barU}) here, using the identity
\begin{equation} \label{idgamma}
     (-1)^{j''}\binom{j'+l}{l}\gamma_{j',j''+j|j}
        - (-1)^{j}\binom{j''+j}{j}\gamma_{j'',j'+l|l}
    = (-1)^l \binom{j+l}{l}\tilde\gamma_{j''+j,j'+l|j+l} \epc
\end{equation}
where
\begin{align} \label{tildegamma}
     \tilde \gamma_{j',j''|l}
        = \biggl(\binom{j'}{l}-(-1)^l\binom{j''}{l}\biggr)\gamma_{j'+j''-l}
\end{align}
and doing some algebra, we come back to our previous compatibility condition
(\ref{eqcbarc}).

Hence, either of the two equations (\ref{cU}) or (\ref{cbarU}) can be used to
determine the coefficients $c_{j, j'}$. Their high-temperature expansion is
of the form
\begin{equation} \label{tempc2}
     c_{j,j'} = \sum\limits_{k>j+j'}\be^k c_{k|j,j'} \epp
\end{equation}
Substituting this together with (\ref{tempc'}), (\ref{U}) into equation
(\ref{cU}) we obtain
\begin{equation} \label{tempc2c}
     (q^{\a}-q^{-\a}) c_{k|j,l}
        = \binom{j+l}{l} c'_{k|j+l}
      -\sum\limits_{k'=j+1}^{k-1} \sum\limits_{j'=0}^{k'-j-1}
       \sum\limits_{j''=0}^{k-k'-1} c_{k'|j,j'} c'_{k-k'|j''} \gamma_{j',j''|l} \epp
\end{equation}
Again, like in equation (\ref{tempeqcc'}), the coefficients $c_{k|j,l}$ are
completely determined by the previous coefficients $c_{k'|j',l'}$ with $k'<k$.
And again we obtain an iterative calculation scheme which is very efficient and allows
us to quickly calculate high-order terms in the high-temperature expansion
(\ref{tempc2}).

\section{Representation of $G$, $\overline G$ and $\om$ in terms of the master
function}
Since we have the formulae (\ref{G}), (\ref{barG}) and (\ref{om}) which
relate the functions $G,\overline G$ and $\om$, respectively, to the resolvent
$R$ given by (\ref{solR}), it is merely a technical problem to express these
three functions in terms of $\Phi(\z,\z'|\kappa,\kappa';\al)$. An important point
here is that it turns out to be possible to get rid of all integrations.
\begin{lem} \label{lem:omphi}
The functions $G,\overline G$ and $\om$ are expressed in terms of the function
$\Phi$ as follows
\begin{align}
     G(\z,\z'|\kappa,\kappa';\al)
        = \Delta_{\z}H_{\z'}\Phi(\z,\z'|\kappa,\kappa';\al) \epc \label{GPhi} \\
     \overline G(\z,\z'|\kappa,\kappa';\al)
        = - \frac{1}{2\pi i}H_{\z}\Delta_{\z'}\Phi(\z,\z'|\kappa,\kappa';\al) \epc
        \label{barGPhi} \\
     \frac14 \om(\z,\z'|\kappa,\kappa';\al)
        = H_{\z}H_{\z'}\Phi(\z,\z'|\kappa,\kappa';\al) \epc \label{omPhi}
\end{align}
where the operators $\Delta_{\z}$ and $H_{\z}$ are defined in (\ref{deltas}) and
(\ref{H}), respectively.
\end{lem}

The proof of this Lemma is deferred to Appendix \ref{app:omphi}.

The master functions $\Phi$ has the symmetry
\begin{equation}
     \Phi (\z, \z'|\k, \k', \a) = \Phi (\z', \z|\k, \k', - \a) 
\end{equation}
which follows from (\ref{masterPhi}), (\ref{masterrelPhi1a}) and (\ref{masterrelPhi2a}).
Due to our Lemmas \ref{lem:R}, \ref{lem:omphi} this symmetry carries over to $R$, $\om$,
$G$ and $\overline G$,
\begin{align}
     & R (\z, \z'|\k, \k', \a) = R (\z', \z|\k, \k', - \a) \epc \\
     & \om (\z, \z'|\k, \k', \a) = \om (\z', \z|\k, \k', - \a) \epc \notag \\
     & \overline{G} (\z, \z'|\k, \k', \a) = - \frac{1}{2 \p i} G (\z', \z|\k, \k', - \a)
         \notag \epp
\end{align}

A few comments are in order here. First, the expression (\ref{omPhi})
for our main function $\om$ is rather appealing as compared to the original
expression (\ref{om}). We could get rid of all unpleasant integrations that
make the analysis hard. Second, we derived our formulae in the temperature
case. Yet, (\ref{solR}), (\ref{GPhi}), (\ref{barGPhi}) and (\ref{omPhi}) are
valid in the general inhomogeneous case for finite $N$. This means that we can
use our formulae for the future analysis of the CFT scaling limit.

As we pointed out above our master function $\Phi$ of two spectral parameters
encodes the complete information about various functions appearing in the
description of the physical part of the correlation functions of the XXZ chain.
This function $\Phi$ can be characterized by either of the relations
(\ref{masterrelPhi1}), (\ref{masterrelPhi2}), (\ref{masterrelPhi1a}) or
(\ref{masterrelPhi2a}) which ensure the cancellation of singularities inside
the integration contour $\gamma$. These relations make sense in the general
case as well, when the lattice is finite in Matsubara direction. In that
case they mean that the residues at all poles corresponding to the Bethe roots
must vanish. In the temperature case we could use the high-temperature expansion
(\ref{phi2expan}) and determine the coefficients $c_{j,j'}$. This procedure
can be efficiently implemented on a computer. We hope that it will be possible
in the future to find a generalization of this procedure to the CFT scaling limit,
where relations like (\ref{masterrelPhi1}), (\ref{masterrelPhi2}) or
(\ref{masterrelPhi1a}), (\ref{masterrelPhi2a}) may turn into a Riemann-Hilbert problem.

We also believe that it would be interesting to understand the deeper meaning
of the operator $H$ defined in (\ref{H}).

\begin{ack}
The authors are indebted to M.~Jimbo, Kh.~Nirov and F.~Smirnov for helpful
discussions. HB is grateful to the Volkswagen Foundation for financial support.
\end{ack}

\vfill

{\appendix
\section{Proof of Lemma \ref{lem:R}} \label{app:proofresolvent}
The statement of the Lemma can be verified rather directly. First we write
(\ref{eqR}) in the following way,
\begin{equation} \label{eqRa}
     R(\z,\z')=K_{\al}(\z/\z')+(K_{\al}\star R)(\z,\z') \epc
\end{equation}
where we omit the other arguments of the resolvent for simplicity. We
will also omit such arguments in all other functions during this proof.

Let us substitute the right hand side of (\ref{solR}) into the right hand
side of (\ref{eqRa}) and use (\ref{masterrelPhi1}). Then
\begin{align*}
     & K_{\a}(\z/\z')
        +\int_{\gamma}\frac{\rd \eta^2}{\eta^2}
       K_{\al}(\z,\eta)\frac{R(\eta,\z')}{\rho(\eta)(1+\mathfrak{a}(\eta))}\\
     & = K_{\a}(\z/\z')
         -\int_{\gamma}\frac{\rd \eta^2}{\eta^2}
     K_{\al}(\z,\eta)\bigl(-\frac{\Phi'(\eta,\tau)}{\Delta_{\eta}\Phi(\eta,\tau)}
                           +r(\eta,\tau)\bigr)
         \frac{1}{2\pi i}\Delta_{\eta}\Delta_{\z'}\Phi(\eta,\z')\\
     & = K_{\a}(\z/\z')
         +\int_{\gamma}\frac{\rd \eta^2}{\eta^2}
     K_{\al}(\z,\eta)\bigl(\frac{\Phi'(\eta,\tau)}{\Delta_{\eta}\Phi(\eta,\tau)}
                           -r(\eta,\tau)\bigr)
         \frac{1}{2\pi i}\Delta_{\eta}(\Phi(\eta,q\z')-\Phi(\eta,q^{-1}\z'))\\
     & = K_{\a}(\z/\z')
         +\int_{\gamma}\frac{\rd \eta^2}{\eta^2} K_{\al}(\z,\eta)
     \biggl(\Bigl(\frac{\Phi'(\eta,q\z')}{\Delta_{\eta}\Phi(\eta,q\z')}
     - r(\eta,q\z')\Bigr) \frac{1}{2\pi i}\Delta_{\eta}\Phi(\eta,q\z')\\
     & \mspace{234.mu}
       - \Bigl(\frac{\Phi'(\eta,q^{-1}\z')}{\Delta_{\eta}\Phi(\eta,q^{-1}\z')}
       - r(\eta,q^{-1}\z')\Bigr)
         \frac{1}{2\pi i}\Delta_{\eta}\Phi(\eta,q^{-1}\z')\biggr) \epc
\end{align*}
where we used the arbitrariness of the parameter $\tau$ and choose either
$\tau=q\z'$ or $\tau=q^{-1}\z'$. We further rewrite the last line as
\begin{align*}
     & K_{\a}(\z/\z')
        +\int_{\gamma}\frac{\rd \eta^2}{\eta^2}K_{\al}(\z,\eta)\frac{1}{2\pi i}\Delta_{\z'}
     \biggl(\Phi'(\eta,\z')-r(\eta,\z')\Delta_{\eta}\Phi(\eta,\z')\biggr) \\
     & = K_{\a}(\z/\z')+\frac{1}{2\pi i}\int_{\gamma}\frac{\rd \eta^2}{\eta^2}
         K_{\al}(\z,\eta)\Delta_{\z'}\Phi'(\eta,\z') \\
     & = K_{\a}(\z/\z')-\frac{1}{2\pi i}\Delta_{\z'}\Delta_{\z}\Phi'(\z,\z')
       = \frac{1}{2\pi i}\Delta_{\z}\psi(\z/\z',\a)
        -\frac{1}{2\pi i}\Delta_{\z}\Delta_{\z'}\Phi'(\z,\z') \\
     & =-\Delta_{\z}\Delta_{\z'}\Bigl(\frac{1}{2\pi i}\Phi'(\z,\z')
        +\frac{1}{2\pi i}\Delta_{\z}^{-1}\psi(\z/\z',\a)\Bigr)
       =-\frac{1}{2\pi i}\Delta_{\z}\Delta_{\z'}\Phi(\z,\z') \epc
\end{align*}
where we deformed the contour in such a way that it does not contain the essential
singularity $\eta=1$, but the two poles coming from the kernel $K_{\al}(\z,\eta)$
instead. One also has to verify that the contribution of the boundary terms is equal
to 0. This can be done in a similar way as it was explained in \cite{BoGo10}.

\section{Proof of Lemma \ref{lem:omphi}}
\label{app:omphi}
(i) Let us start with (\ref{GPhi}). In accordance with the formula (\ref{G}) we
have (again we keep only spectral parameters in arguments and omit all other
arguments)
\begin{align*}
     & G(\z,\z') = f_\mathrm{right}(\z,\z')+(R\star f_\mathrm{right})(\z,\z') \epc
                   \quad \\
     & f_\mathrm{right}(\z,\z') 
                                = \psi(q^{-1}\z/\z')-\rho(\z')\psi(\z/\z') \epp
\end{align*}
Hence, the integral on the right hand side is
$$
     (R\star f_\mathrm{right})(\z,\z')
        = \int_{\gamma}\frac{\rd \eta^2}{\eta^2}
      \frac{R(\z,\eta)}{\rho(\eta)(1+\mathfrak{a}(\eta))}
      f_\mathrm{right}(\eta,\z')=I_1-\rho(\z')I_2
$$
where both spectral parameters $\z$ and $\z'$ are inside the integration contour
$\gamma$ and
$$
     I_1 = \int_{\gamma}\frac{\rd \eta^2}{\eta^2}
           \frac{R(\z,\eta)}{\rho(\eta)(1+\mathfrak{a}(\eta))}\psi(q^{-1}\eta/\z')
       \epc \qd
     I_2 = \int_{\gamma}\frac{\rd \eta^2}{\eta^2}
           \frac{R(\z,\eta)}{\rho(\eta)(1+\mathfrak{a}(\eta))}\psi(\eta/\z') \epp
$$

We treat the integral $I_1$ in a similar way as in the proof of Lemma \ref{lem:R},
\begin{align*}
     I_1 & = -\int_{\gamma}\frac{\rd \eta^2}{\eta^2}\frac{1}{2\pi i}
              \Delta_{\z}\Delta_{\eta}\Phi(\z,\eta)
              \Bigl(-\frac{\Phi'(\tau,\eta)}{\Delta_{\eta}\Phi(\tau,\eta)}
                +r'(\tau,\eta)\Bigr)\psi(q^{-1}\eta/\z') \\
           & = -\int_{\gamma}\frac{\rd \eta^2}{\eta^2}\frac{1}{2\pi i}
                \Bigl(\Delta_{\eta}\Phi(q\z,\eta)
              -\Delta_{\eta}\Phi(q^{-1}\z,\eta)\Bigr)
                \Bigl(-\frac{\Phi'(\tau,\eta)}{\Delta_{\eta}\Phi(\tau,\eta)}
              + r'(\tau,\eta)\Bigr)\psi(q^{-1}\eta/\z') \\
           & = -\int_{\gamma}\frac{\rd \eta^2}{\eta^2}\frac{1}{2\pi i}
                \biggl(\Delta_{\eta}\Phi(q\z,\eta)
        \Bigl(-\frac{\Phi'(q\z,\eta)}{\Delta_{\eta}\Phi(q\z,\eta)}
              +r'(q\z,\eta)\Bigr) \\
           & \mspace{154.mu}
             - \Delta_{\eta}\Phi(q^{-1}\z,\eta)
            \Bigl(-\frac{\Phi'(q^{-1}\z,\eta)}{\Delta_{\eta}\Phi(q^{-1}\z,\eta)}
              +r'(q^{-1}\z,\eta)\Bigr) \biggr)\psi(q^{-1}\eta/\z') \epp
\end{align*}
The integrals containing the remainder function $r'$ do not contribute, because
all singularities are outside the integration contour. We conclude that
$$
     I_1 = \int_{\gamma}\frac{\rd \eta^2}{\eta^2}\frac{1}{2\pi i}
           \Bigl(\Phi'(q\z,\eta)-\Phi'(q^{-1}\z,\eta)\Bigr)\psi(q^{-1}\eta/\z')
     = \Delta_{\z}\Phi'(\z,q\z') \epp
$$

In the second integral $I_2$ we have to take into account the residue originating
from $\psi(\eta/\z')$:
$$
     I_2 = -2\pi i\frac{R(\z,\z')}{\rho(\z')(1+\mathfrak{a}(\z'))}
           +\int_{\gamma'}\frac{\rd \eta^2}{\eta^2}
          \frac{R(\z,\eta)}{\rho(\eta)(1+\mathfrak{a}(\eta))}\psi(\eta/\z') \epc
$$
where the contour $\gamma'$ does not contain the point $\z'$. The corresponding
integral can be calculated in the same manner as $I_1$. The result is
$$
     I_2 = -2\pi i\frac{R(\z,\z')}{\rho(\z')(1+\mathfrak{a}(\z'))}
           +\Delta_{\z}\Phi'(\z,\z') \epp
$$

Altogether
\begin{align*}
     G(\z,&\z') =\psi(q^{-1}\z/\z')-\rho(\z')\psi(\z/\z')+I_1-\rho(\z_1)I_2 \\
        & = \psi(q^{-1}\z/\z')-\rho(\z')\psi(\z/\z')+\Delta_{\z}\Phi'(\z,q\z')
        +2\pi i\frac{R(\z,\z')}{1+\mathfrak{a}(\z')}
        -\rho(\z')\Delta_{\z}\Phi'(\z,\z') \\
        & = \Delta_{\z}\Phi(\z,q\z')-\rho(\z')\Delta_{\z}\Phi(\z,\z')
       -\frac{1}{1+\mathfrak{a}(\z')} \Delta_{\z}\Delta_{\z'}\Phi(\z,\z') \\
        & = \Delta_{\z}\Bigl(\delta^-_{\z'}
        -\frac{1}{1+\mathfrak{a}(\z')}\Delta_{\z'}\Bigl)\Phi(\z,\z')
      = \Delta_{\z}H_{\z'}\Phi(\z,\z') \epp
\end{align*}

(ii) The proof of the formula (\ref{barGPhi}) proceeds similarly to the above proof
of (\ref{GPhi}). We end up with
$$
     \overline G(\z,\z')
        = -\frac{1}{2\pi i}\Bigl(\delta^-_{\z}
      -\frac{1}{1+\mathfrak{a}(\z)}\Delta_{\z}\Bigl)\Delta_{\z'}\Phi(\z,\z')
    = -\frac{1}{2\pi i}H_{\z}\Delta_{\z'}\Phi(\z,\z') \epp
$$

(iii) Finally we prove (\ref{omPhi}) which is the main result of this paper. The
function $\om$ is given by (\ref{om})
\begin{equation} \label{om1}
     \frac{1}4\om(\z,\z')
        = (\overline G\star f_\mathrm{right})(\z,\z')
      + \delta^-_{\z}\delta^-_{\z'}\Delta_{\z}^{-1}\psi(\z/\z') \epp
\end{equation}
After inserting (\ref{barGPhi}) the first term on the right hand takes the form
\begin{multline} \label{I+0-}
     - \int_{\gamma}\frac{\rd \eta^2}{\eta^2} \frac{1}{2\pi i}
                    \frac{1}{\rho(\eta)(1+\mathfrak{a}(\eta))}
            H_{\z}\Delta_{\eta}\Phi(\z,\eta) f_\mathrm{right}(\eta,\z') = \\
        -\frac{1}{1+\bar{\mathfrak{a}}(\z)}I_1^{(+)}
    -\frac{1}{1+\mathfrak{a}(\z)}I_1^{(-)}
    +\rho(\z)I_1^{(0)} +\frac{\rho(\z')}{1+\bar{\mathfrak{a}}(\z)}I_2^{(+)}
    +\frac{\rho(\z')}{1+\mathfrak{a}(\z)}I_2^{(-)}-\rho(\z)\rho(\z')I_2^{(0)} \epc
\end{multline}
where
\begin{align*}
     & I_1^{(\pm )} = \frac1{2\pi i}\int_{\gamma}\frac{\rd \eta^2}{\eta^2}
                      \frac{1}{\rho(\eta)(1+\mathfrak{a}(\eta))}
              \Delta_{\eta}\Phi(q^{\pm 1}\z,\eta)\psi(q^{-1}\eta/\z') \epc \\
     & I_2^{(\pm )} = \frac1{2\pi i}\int_{\gamma}\frac{\rd \eta^2}{\eta^2}
                      \frac{1}{\rho(\eta)(1+\mathfrak{a}(\eta))}
              \Delta_{\eta}\Phi(q^{\pm 1}\z,\eta)\psi(\eta/\z') \epc \\
     & I_1^{(0)} = \frac1{2\pi i}\int_{\gamma}\frac{\rd \eta^2}{\eta^2}
                   \frac{1}{\rho(\eta)(1+\mathfrak{a}(\eta))}
           \Delta_{\eta}\Phi(\z,\eta)\psi(q^{-1}\eta/\z') \epc \\
     & I_2^{(0)} = \frac1{2\pi i}\int_{\gamma}\frac{\rd \eta^2}{\eta^2}
                   \frac{1}{\rho(\eta)(1+\mathfrak{a}(\eta))}
           \Delta_{\eta}\Phi(\z,\eta)\psi(\eta/\z') \epp
\end{align*}

In order to calculate all these integrals we proceed in a similar way as before
\begin{multline}
     I_1^{(\pm )} =
        \frac1{2\pi i}\int_{\gamma}\frac{\rd \eta^2}{\eta^2}
    \Bigl(-\frac{\Phi'(q^{\pm 1}\z,\eta)}{\Delta_{\eta}\Phi(q^{\pm 1}\z,\eta)}
          +r(q^{\pm 1}\z,\eta)\Bigr)
    \Delta_{\eta}\Phi(q^{\pm 1}\z,\eta)\psi(q^{-1}\eta/\z')\label{resI1+-} \\
        = -\frac1{2\pi i}\int_{\gamma}\frac{\rd \eta^2}{\eta^2}
       \Phi'(q^{\pm 1}\z,\eta)\psi(q^{-1}\eta/\z')
    =-\Phi'(q^{\pm 1}\z,q\z') \epp
\end{multline}
\begin{multline*}
     I_2^{(\pm )} =
        \frac1{2\pi i}\int_{\gamma}\frac{\rd \eta^2}{\eta^2}
        \Bigl(-\frac{\Phi'(q^{\pm 1}\z,\eta)}{\Delta_{\eta}\Phi(q^{\pm 1}\z,\eta)}
          +r(q^{\pm 1}\z,\eta)\Bigr)
        \Delta_{\eta}\Phi(q^{\pm 1}\z,\eta)\psi(\eta/\z') \\
        = \frac1{2\pi i}\int_{\gamma}\frac{\rd \eta^2}{\eta^2}
      \Bigl(-\Phi'(q^{\pm 1}\z,\eta)
            +r(q^{\pm 1}\z,\eta)\Delta_{\eta}\Phi(q^{\pm 1}\z,\eta)\Bigr)
        \psi(\eta/\z') \epp
\end{multline*}
Since $\z'$ is inside the contour one can see that the first term inside the
bracket does not contribute, and we have
\begin{multline} \label{resI2+-}
     I_2^{(\pm )}
        = \frac1{2\pi i}\int_{\gamma}\frac{\rd \eta^2}{\eta^2}
      r(q^{\pm 1}\z,\eta)\Delta_{\eta}\Phi(q^{\pm 1}\z,\eta) \psi(\eta/\z') \\
    = -r(q^{\pm 1}\z,\z')\Delta_{\z'}\Phi(q^{\pm 1}\z,\z')
        = -\Phi'(q^{\pm 1}\z,\z')
      -\frac{\Delta_{\z'}\Phi(q^{\pm 1}\z,\z')}{\rho(\z')(1+\mathfrak{a}(\z'))} \epp
\end{multline}
Furthermore,
\begin{align}
     I_1^{(0)} & =
        \frac1{2\pi i}\int_{\gamma}\frac{\rd \eta^2}{\eta^2}
    \Bigl(-\frac{\Phi'(\z,\eta)}{\Delta_{\eta}\Phi(\z,\eta)}+r(\z,\eta)\Bigr)
        \Delta_{\eta}\Phi(\z,\eta)\psi(q^{-1}\eta/\z')\label{resI10}\\
        & = -\Phi'(\z,q\z')-r(\z,\z)\psi(q^{-1}\z/\z') \notag \\
    & = -\Phi'(\z,q\z')-\biggl(\frac1{\rho(\z)(1+\mathfrak{a}(\z))}
        +\lim_{\z''\to\z}\frac{\Phi'(\z,\z'')}{\Delta_{\z''}\Phi(\z,\z'')}\biggr)
        \psi(q^{-1}\z/\z') \notag \\
        & = -\Phi'(\z,q\z')
        -\frac1{\rho(\z)(1+\mathfrak{a}(\z))}\psi(q^{-1}\z/\z') \notag \epc
\end{align}
since the term with $\lim_{\z''\to\z}$ vanishes. Finally
\begin{align}
     I_2^{(0)} & =
        \frac1{2\pi i}\int_{\gamma}\frac{\rd \eta^2}{\eta^2}
    \Bigl(-\frac{\Phi'(\z,\eta)}{\Delta_{\eta}\Phi(\z,\eta)}+r(\z,\eta)\Bigr)
        \Delta_{\eta}\Phi(\z,\eta)\psi(\eta/\z')\label{resI20}\\
        & = -r(\z,\z')\Delta_{\z'}\Phi(\z,\z')-r(\z,\z)\psi(\z/\z') \notag \\
        & = -\Phi'(\z,\z')
        -\frac{\Delta_{\z'}\Phi(\z,\z')}{\rho(\z')(1+\mathfrak{a}(\z'))}
        -\frac1{\rho(\z)(1+\mathfrak{a}(\z))}\psi(\z/\z')\notag \epp
\end{align}

If we now substitute (\ref{resI1+-}), (\ref{resI2+-}), (\ref{resI10}), (\ref{resI20})
into the expression (\ref{I+0-}), combine with the last term at the right hand
side of (\ref{om1}) and do some algebra, we come to the final result (\ref{omPhi}).

\section{Coefficients in the high-temperature expansion of the function
$\phi(\la|\kappa,\kappa')$} \label{app:htphi}
As we mentioned in Section \ref{subsec:sigmaphi}, we can compute the
coefficients $c_j$ of the expansion (\ref{expanphi}) by solving the
functional equation (\ref{Phiauxfun}) and by using the corresponding
high-temperature expansion (\ref{tempc}). Unfortunately, the resulting
formulae for the coefficients $c_{k|j}$ are not simple, not even at lower
orders. There is a better object though, for which the coefficients of
the high-temperature expansion are much simpler and from which the
coefficients $c_{k|j}$ can be obtained.

It might seem that the function $A$ considered in Section \ref{subsec:sigmaphi}
is a candidate to be such simple function. Unfortunately, $A$ is not well-defined
in the temperature case when the limit $N\to\infty$ is performed. Therefore we
need another function $\tilde A$ which exists in this limit. In order to define
it we first introduce a function $\tilde\phi$ by changing the normalization of $\phi$,
\begin{equation}
     \Phi(\z|\kappa,\kappa')
        = \z^{-\kappa'+\kappa} \tilde \phi(\la|\kappa,\kappa') \epc
	  \quad \la=\log{\z} \epp
\end{equation}

Due to (\ref{expanphi}) the function $\tilde\phi$ has an expansion of the form
\begin{equation} \label{expanphitilde}
     \tilde \phi(\la|\kappa,\kappa')
        = 1+\sum\limits_{j\ge 0} \tilde c_j\cth^{(j)}(\la) \epc
          \qd \tilde c_j=\sum\limits_{k>j} {\be}^k \tilde c_{k|j} \epp
\end{equation}
It can be obtained from the expansion (\ref{expanphi}) by taking into account that $\Phi_0
= \bigl(\tilde \phi(\infty|\kappa,\kappa') + \tilde \phi(-\infty|\kappa,\kappa')\bigr)/2$.
Then there is a unique solution $\tilde A(\la, \k)$ of the functional
equation
\begin{equation}
     \tilde \phi(\la|\kappa,\kappa') = \frac{\tilde A(\la, \k')}{\tilde A(\la,\k)}
\end{equation}
in form of a high-temperature series
\begin{equation} \label{Aexpan}
    \tilde A(\la ,\kappa)
       = \exp \biggl\{ \sum\limits_{j\ge 0}a_j \coth^{(j)}(\la) \biggr\} \epc \quad
         a_j=\sum\limits_{k>j} {\be}^k a_{k|j} \epc
\end{equation}
with coefficients $a_{k|j}$ which depend only on $\kappa$. These coefficients
are comparatively simple.

Here are the first few of them,
\begin{align*}
&a_{1|0}=-(q-q^{-1})/v,\quad v:= q^{2\kappa}+1 \epc \\
&a_{2|0}=-(q^2-q^{-2})(v-1)(v-2)/v^3 \epc \\
&a_{2|1}=-(q-q^{-1})^2(v-1)/(2v^2) \epc \\
&a_{3|0}=-2(q-q^{-1})(v-1)(v-2) \bigl((q^2+q^{-2})(v^2-6v+6)+4v^2-12v+12\bigr)/(3v^5)
          \epc \\
&a_{3|1}=-(q-q^{-1})(q^2-q^{-2})(v-1)(v-2)^2/(2v^4) \epc \\
& a_{3|2}=-(q-q^{-1})^3(v-1)(v-2)/(12v^3) \epc \\
&a_{4|0}=-(q^2-q^{-2})(v-1)(v-2) \bigl( (q^2+q^{-2}) (v^4-18v^3+78v^2-120v+60) \\
& \mspace{324.mu} +8(v^4-9v^3+24v^2-30v+15)\bigr)/(3v^7)\epc \\
&a_{4|1}=-(q-q^{-1})^2(v-1) \bigl( (q^2+q^{-2})(11v^4-138v^3+498 v^2-720v+360) \\
& \mspace{243.mu} +2(19v^4-186v^3+546 v^2-720v+360) \bigr)/(36v^6) \epc \\
&a_{4|2}=-(q-q^{-1})^2(q^2-q^{-2})(v-1)(v-2)(v^2-6v+6)/(12v^5) \epc \\
&a_{4|3}=-(q-q^{-1})^4(v-1)(v^2-6v+6)/(144v^4) \epp
\end{align*}

The explicit coefficients $c_{k|j}$ can now be obtained following the above
prescription in opposite direction.

\section{High-temperature expansion of the function $\Phi(\z|\kappa,\kappa';\al)$}
\label{app:htgenphi}
Following the above scheme, one can obtain the coefficients $c_j$ up to any
order in $\be$. In this sense we shall imply now that all these coefficients
are already known. The next step is to calculate the coefficients $c'_j$ from
(\ref{expanphi'}) which determine the function $\Phi(\z|\kappa,\kappa';\al)$ given by
(\ref{genasPhi}). To this end we can simply solve equation (\ref{eqcc'}) with
respect to the $c'_{k|j}$. Here are the first coefficients obtained this way
\begin{align*}
&c'_{1|0}/\chi = c_{1|0} ,\quad
            \chi:=\frac{q^{\a}-q^{-\a}}{q^{\kappa'-\kappa}-q^{-\kappa'+\kappa}} \epc \\
&c'_{2|0}/\chi = c_{2|0} \\
&\qd + \frac{(q+q^{-1}) \bigl(q^{(\kappa'-\kappa+\a)/2}-q^{-(\kappa'-\kappa+\a)/2}\bigr)
\bigl(q^{(\kappa'-\kappa-\a)/2}-q^{-(\kappa'-\kappa-\a)/2}\bigr)}
{(q-q^{-1})(q^{\kappa'-\kappa}-q^{-\kappa'+\kappa})} c_{1|0}^2 \epc \\
\displaybreak[0]
&c'_{2|1}/\chi = c_{2|1} \epc \\
&c'_{3,0}/\chi = c_{3|0} \\
&\qd +2 \frac{(q+q^{-1})
\bigl(q^{(\kappa'-\kappa+\a)/2}-q^{-(\kappa'-\kappa+\a)/2}\bigr)
\bigl(q^{(\kappa'-\kappa-\a)/2}-q^{-(\kappa'-\kappa-\a)/2}\bigr)}
{(q-q^{-1})(q^{\kappa'-\kappa}-q^{-\kappa'+\kappa})}
c_{1|0}c_{2|0} \epc \\
&\qd + \frac{(q+q^{-1})^2
\bigl(q^{(\kappa'-\kappa+\a)/2}-q^{-(\kappa'-\kappa+\a)/2}\bigr)^2
\bigl(q^{(\kappa'-\kappa-\a)/2}-q^{-(\kappa'-\kappa-\a)/2}\bigr)^2}
{(q-q^{-1})^2(q^{\kappa'-\kappa}-q^{-\kappa'+\kappa})^2} c_{1|0}^3 \epc \\
&c'_{3|1}/\chi = c_{3|1}\\
&\qd + \frac{(q+q^{-1})
\bigl(q^{(\kappa'-\kappa+\a)/2}-q^{-(\kappa'-\kappa+\a)/2}\bigr)
\bigl(q^{(\kappa'-\kappa-\a)/2}-q^{-(\kappa'-\kappa-\a)/2}\bigr)}
{(q-q^{-1})(q^{\kappa'-\kappa}-q^{-\kappa'+\kappa})} c_{1|0}c_{2|1} \epc \\
&c'_{3|2}/\chi = c_{3|2} \epp
\end{align*}

\section{High-temperature coefficients for $\Phi(\z,\z'|\kappa,\kappa';\a)$}
Here we show some coefficients $c_{j,j'}, c_{k|j,j'}$ that determine the function
$\Phi(\z,\z'|\kappa,\kappa';\a)$ via the expansions (\ref{phi2expan}) and
(\ref{tempc2}). Let us start with two useful observations:

\begin{rem} \label{rem:outercoeff}
Analyzing the equations (\ref{cU}), (\ref{cbarU}), we observe the following
\begin{align}
     c_{j,0}=-\frac{\bar c'_{j}}{q^{\a}-q^{-\a}} \epc \quad
     c_{0,j}=\frac{c'_{j}}{q^{\a}-q^{-\a}} \label{cj0} \epp
\end{align}
This can also be deduced from the limiting relations (\ref{limitPhi}). As in
the Remark above, these two relations hold to all orders in the high-temperature
expansions.
\end{rem}

\begin{rem}
The equations (\ref{cU}), (\ref{cbarU}) have the symmetry
\begin{align}
     & c'_j\leftrightarrow-(-1)^j\bar c'_j \epc \label{ctobarc}\\
     & c_{j,j'}\leftrightarrow (-1)^{j+j'}c_{j',j}
       \notag \epp
\end{align}
It is easy to check that the equation (\ref{eqcbarc}) is also explicitly symmetric
under the transformation (\ref{ctobarc}).
\end{rem}

Due to Remark \ref{rem:outercoeff} it is easy to obtain the high-temperature
expansions for $c_{j,0}$ and $c_{0,j}$. We will not show the corresponding
coefficients $c_{k|j,0}$ and $c_{k|0,j}$ here. The other coefficients can
be obtained solving equation (\ref{tempc2c}) order by order. Here are several
examples:
\begin{align*}
&c_{3|1,1}=\frac{2}{q^{\a}-q^{-\a}}\;c'_{3|2} \epc \\
&c_{4|1,1}=\frac{2}{q^{\a}-q^{-\a}}\;c'_{4|2}
           +\frac{2(q+q^{-1})(q^{\a}+q^{-\a})}{(q-q^{-1})
        (q^{\a}-q^{-\a})^2}\;c'_{1|0}c'_{3|2}
       -\frac{(q+q^{-1})(q^{\a}+q^{-\a})}
             {(q-q^{-1})(q^{\a}-q^{-\a})^2}\;{(c'_{2|1})}^2, \\
&c_{4|1,2}=c_{4|2,1}=\frac{3}{q^{\a}-q^{-\a}}\;c'_{4|3} \epp
\end{align*}
All the coefficients $c'_{k|j}$ can be obtained following the previous Appendix.
With the help of a computer one can efficiently obtain higher coefficients $c'_{k|j}$
and $c_{k|j,j'}$ as well by means of the above scheme.
}

\bigskip

\providecommand{\bysame}{\leavevmode\hbox to3em{\hrulefill}\thinspace}
\providecommand{\MR}{\relax\ifhmode\unskip\space\fi MR }
\providecommand{\MRhref}[2]{%
  \href{http://www.ams.org/mathscinet-getitem?mr=#1}{#2}
}
\providecommand{\href}[2]{#2}


\end{document}